\documentclass[sigconf]{cidr-2024}
\usepackage{subfig} 
\usepackage{xcolor}
\usepackage{tabularray}
\usepackage{amsmath}
\UseTblrLibrary{booktabs}
\DeclareMathOperator*{\argmax}{arg\,max}
\usepackage{xspace}

\newcommand{\nopt}{{\sc NoPredTrans}\xspace}
\newcommand{\bj}{{\sc BloomJoin}\xspace}
\newcommand{\yk}{{\sc Yannakakis}\xspace}
\newcommand{\pt}{{\sc PredTrans}\xspace}

\begin{document}

\title{Predicate Transfer: Efficient Pre-Filtering on Multi-Join Queries}

\author{
    Yifei Yang, Hangdong Zhao, Xiangyao Yu, Paraschos Koutris}
\affiliation{%
  \institution{University of Wisconsin-Madison} 
}
\email{yyang673@wisc.edu, {hangdong,yxy,paris}@cs.wisc.edu}



\captionsetup[figure]{font=small}
\captionsetup[table]{font=small}
\captionsetup[lstlisting]{font=small}

\begin{abstract}
This paper presents \textit{predicate transfer}, a novel method that optimizes join performance by pre-filtering tables to reduce the join input sizes.
Predicate transfer generalizes Bloom join, which conducts pre-filtering within a single join operation, to multi-table joins such that the filtering benefits can be significantly increased. Predicate transfer is inspired by the seminal theoretical results by Yannakakis, which uses semi-joins to pre-filter acyclic queries. Predicate transfer generalizes the theoretical results to any join graphs and use Bloom filters to replace semi-joins leading to significant speedup. Evaluation shows predicate transfer can outperform Bloom join by $3.3\times$ on average on TPC-H benchmark. 
\end{abstract}

\maketitle

\section{Introduction}
\label{sec:intro}

Joins constitute a substantial portion of query execution time, and have been studied and optimized for decades, in topics including binary joins (with a main focus on hash joins)~\cite{bj1, bj2, bj3, bj4}, join ordering in multi-way joins~\cite{jo1, jo2, jo3, jo4, jo5}, and recent emerging worst-case optimal join algorithms~\cite{wcoj1, wcoj2, wcoj3, wcoj4}.
One effective principle for enhancing join performance is to minimize the join input sizes by pre-filtering rows that will not appear in the join result. 
Predicate pushdown~\cite{pred-push1, pred-push2, pred-push3, pred-push4, pred-push5, pred-push6} exemplifies this principle by applying local predicates on a table before executing any join operation. 


The Bloom join~\cite{ramesh2009optimizing, bloomjoin2, bloomjoin3} extends this principle beyond a single table. In the Bloom join, a Bloom filter is constructed using the join key in one table, and sent to the other table to filter out rows that do not pass the filter---these rows do not match any keys in the first table and will not participate in the join. 
The Bloom join can effectively reduce the join input sizes thereby reducing the query runtime. However, existing Bloom join solutions can perform such pre-filtering only within a single join operation. 

In this paper, we further generalize the pre-filtering principle 
across multiple joins. Namely, we use predicates on individual tables to pre-filter multiple other tables in the query, further reducing the join input sizes. 
We call this new technique \textit{\textbf{predicate transfer}}. 
A predicate on one table $T_1$ can be transferred (e.g., in the form of a Bloom filter) to a table $T_2$ that joins with $T_1$. $T_2$ can apply the predicate and further transfer it to table $T_3$ that joins with $T_2$  (but $T_1$ does not necessarily join with $T_3$). The transfer process can propagate further such that the original predicate can filter multiple other tables (e.g, $T_2$, $T_3$, etc.).
The conventional Bloom join is a special case of the more generalized predicate transfer---a Bloom join is a one-hop predicate transfer. 

The idea of predicate transfer is inspired by the seminal paper~\cite{Yannakakis1981} by 
Yannakakis. For an acyclic query
that equi-joins multiple tables, the Yannakakis algorithm achieves the theoretically maximum pre-filtering selectivity by adding 
an additional semi-join phase prior to the actual joins, which filters a table by semi-joining it with other tables. 
The process filters one table at a time following the tree structure of the query until every predicate is spread across all joining tables. 

For all its theoretical elegance, the Yannakakis algorithm has not yet made its way into modern database engines. The main obstacles are the costly hash table accesses and high memory consumption 
in the semi-join phase. Predicate transfer aims to address these practical limitations. It significantly reduces the overhead of semi-joins by passing succinct data structures like Bloom filters. Although predicate transfer no longer achieves the theoretically maximum filtering selectivity, it achieves much higher performance overall. 


In the rest of the paper, we first describe the background and related work of predicate transfer in Section~\ref{sec:back}, with a focus on the Bloom join and Yannakakis algorithm. We then describe the design space of predicate transfer in detail, and our current heuristics in different design dimensions in Section~\ref{sec:pred}. We report preliminary performance evaluations on TPC-H~\cite{tpch} in Section~\ref{sec:eval}, which shows that on average predicate transfer can outperform Bloom join by 3.3$\times$ (up to $61\times$) and the Yannakakis algorithm by 4.8$\times$ (up to $47\times$) respectively. 
Finally, Section~\ref{sec:conclusion} concludes the paper and discusses future work.

\section{Background and Related Work}
\label{sec:back}

This section presents the background and related work in Bloom join (Section~\ref{sec:back-bloom}) and the Yannakakis algorithm (Section~\ref{sec:back-yannakakis}). 

\subsection{Bloom join}
\label{sec:back-bloom}

A Bloom filter~\cite{bloom1970space, bf2, bf3, bf4, bf5} is a compact probabilistic data structure that determines whether an element exists in a set. A Bloom filter has no false negative but may have false positives. In a Bloom join of two tables, a Bloom filter is constructed on one table (typically the smaller one) using the join key. The filter is then sent and applied to each row in the other table; if a row does not pass the filter, it matches no row in the first table and should not participate in the join. Since testing a Bloom filter is generally faster than performing a join, Bloom join can speedup query processing, especially when the join is selective. Modern OLAP DBMSs (e.g., Oracle~\cite{oracle-bf}, Redshift~\cite{redshift-bf}, Snowflake~\cite{snowflake-bf}, Databricks~\cite{databricks-bf}) widely adopt Bloom filters to accelerate join execution.

Most existing Bloom join algorithms can be applied to only a single join operation. This means the predicate on one table can only be used to pre-filter rows in the other table it joins with; namely, the predicate is transferred in one-hop and one-direction. Some prior work~\cite{lip17} has extended the idea to datasets with star schemas, allowing all dimension tables to transfer local predicates to the fact table, which outperformed the baseline Bloom join. However, these solutions do not generalize to more complex query plans.

\subsection{Yannakakis algorithm}
\label{sec:back-yannakakis}

The Yannakakis algorithm~\cite{Yannakakis1981} is a classic algorithm that can pre-filter out all rows from tables that do not appear in the final join result, thereby achieving the theoretically maximum filtering selectivity. The algorithm applies to \textit{acyclic} join queries. The acyclicity is more formally termed as \textit{$\alpha$-acyclicity}~\cite{Yannakakis1981}. The algorithm is proven to run in $O(N + \mathsf{OUT})$ time, where $N$ is the size of input relations and $\mathsf{OUT}$ is the query output size. Thus, the Yannakakis algorithm is known to be {\em instance optimal} since 
$N + \mathsf{OUT}$ is the unavoidable time cost of reading the input and enumerating the output for a query.
The algorithm starts by choosing a rooted join tree arbitrarily, and then proceeds with a {\em semi-join phase} and a {\em join phase}. 

\vspace{0.05in}
\noindent
\textbf{Semi-join phase.} The semi-join phase contains 
two passes: the {\em forward pass} and the {\em backward pass}. The forward pass traverses the join tree in a bottom-up fashion. At each vertex, we filter the table by a sequence of semi-joins with its children. A {\em semi-join} of two tables $R$ and $S$ is defined as $R \ltimes S = \Pi_{\mathsf{attr}(R)} (R \bowtie S)$, which effectively removes all tuples in $R$ that do not join with any tuple in $S$. The forward pass stops when the root node is reached. Similarly, the backward pass traverses the join tree in a top-down fashion. At each vertex, the table is filtered by a semi-join with its parent. The backward pass stops when all leaf nodes are reached. It is proven that both passes can be executed in $O(N)$ time and all tuples that will not contribute to the output are removed.

\vspace{0.05in}
\noindent
\textbf{Join phase.} The join phase can join the filtered tables in any order. It is proven that regardless of the chosen join order, the join phase can be executed in $O(\mathsf{OUT})$ time.

As a reflection, the semi-join phase filters all redundant tuples and the join phase executes the join with automatic robustness: it can join the tables in any order without any intermediate table size blow-up over the output size. The algorithm was later extended by Joglekar et al.~\cite{Joglekar2016} to handle aggregations on top of join queries.



\section{Predicate Transfer}
\label{sec:pred}

\begin{figure}
\subfloat[Join Graph]{
	\centering
	\includegraphics[width=0.8\columnwidth]{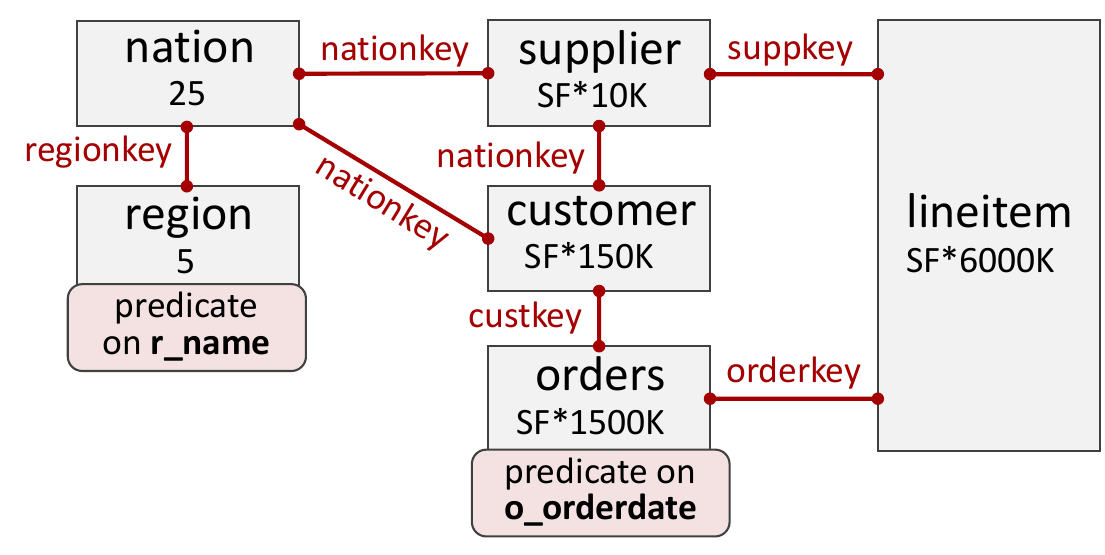}
	\label{fig:join-graph}
} \\
\subfloat[Predicate Transfer Graph]{
	\centering
	\includegraphics[width=0.95\columnwidth]{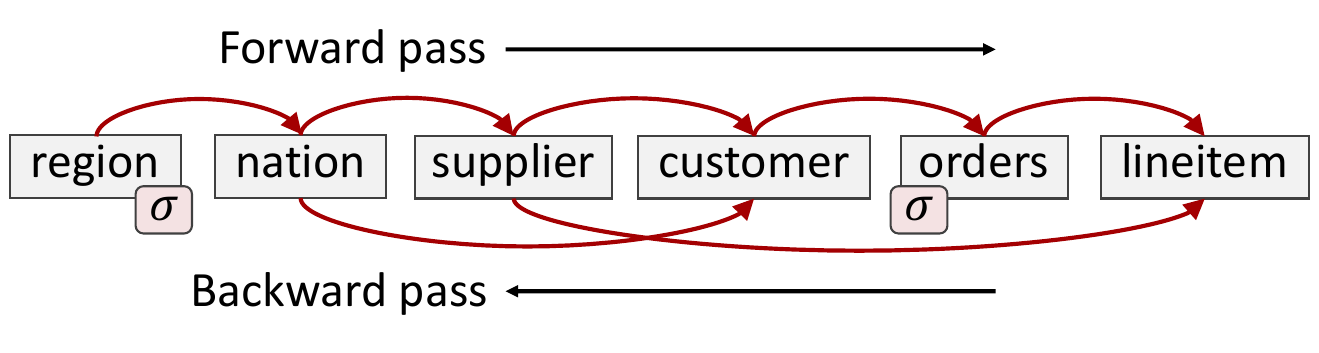}
	\label{fig:pred-trans-graph}
}
\caption{Predicate Transfer for TPC-H Q5.}
\label{fig:q5}
\end{figure}

This section describes the proposed predicate transfer algorithm. We use Query 5 in TPC-H benchmark~\cite{tpch} (Figure~\ref{fig:q5}) as a running example. This query contains six tables, six inner joins, and two predicates on tables \texttt{region} and \texttt{orders} respectively. The discussion assumes equi-join between tables. 


\subsection{Overview}
\label{sec:overview}

Similar to the Yannakakis algorithm, predicate transfer executes a query in two phases. 

\vspace{0.05in}
\noindent
\textbf{Phase 1: Predicate Transfer Phase.} A join graph is constructed for a query, where each vertex is a table and each edge is a join operation. 
A local predicate is constructed as a filter (e.g., a Bloom filter) and be transferred across the join graph. 
The schedule of the predicate transfer phase introduces a large design space, which we discuss in Section~\ref{sec:transfer-phase}. 

\vspace{0.05in}
\noindent
\textbf{Phase 2: Join Phase.}
After the transfer phase finishes, each table has multiple filters, including both local filters and transferred filters. The database can now apply the filters and perform regular joins. The actual inputs of each join will be substantially smaller if the transferred filters are selective. We discuss the join phase in Section~\ref{sec:join-phase}.

\vspace{0.05in}
In the next two subsections, we will describe the design space of these two phases and the heuristics we currently use to implement predicate transfer in our prototype. These heuristics are largely intuition-based and a more thorough theoretical analysis is left for future work.

\subsection{Predicate Transfer Phase}
\label{sec:transfer-phase}

In the rest of this section, we layout the design space of the transfer phase and describe the design choices we adopt in our prototype.

\vspace{0.05in}
\noindent
\textbf{Filter Transformation.}  When transferring a filter across edges that have different join keys, the filter must be transformed. For example, a filter constructed on \texttt{region} can be transferred to \texttt{nation}, but the same filter cannot be directly sent to \texttt{supplier} from \texttt{nation} in the next transfer hop since the join keys do not match. We use the following algorithm to handle the join key mismatch between incoming and outgoing edges on \texttt{nation}. When the incoming filter is received, an empty outgoing filter is created. Then, the columns for both incoming and outgoing join keys in \texttt{nation} are scanned (assuming columnar store; otherwise scan the entire table). Inherent filters of \texttt{nation} are applied during the scan. Then for each row, the incoming join key is used to probe the incoming filter. If a match occurs, the corresponding outgoing join key is added to the outgoing filter. At the end of the scan, the outgoing filter is sent to downstream tables (i.e., \texttt{supplier}). The algorithm is efficient as it requires scanning the join keys only once.

\begin{figure}[h]
    \centering
    \includegraphics[width=\columnwidth]{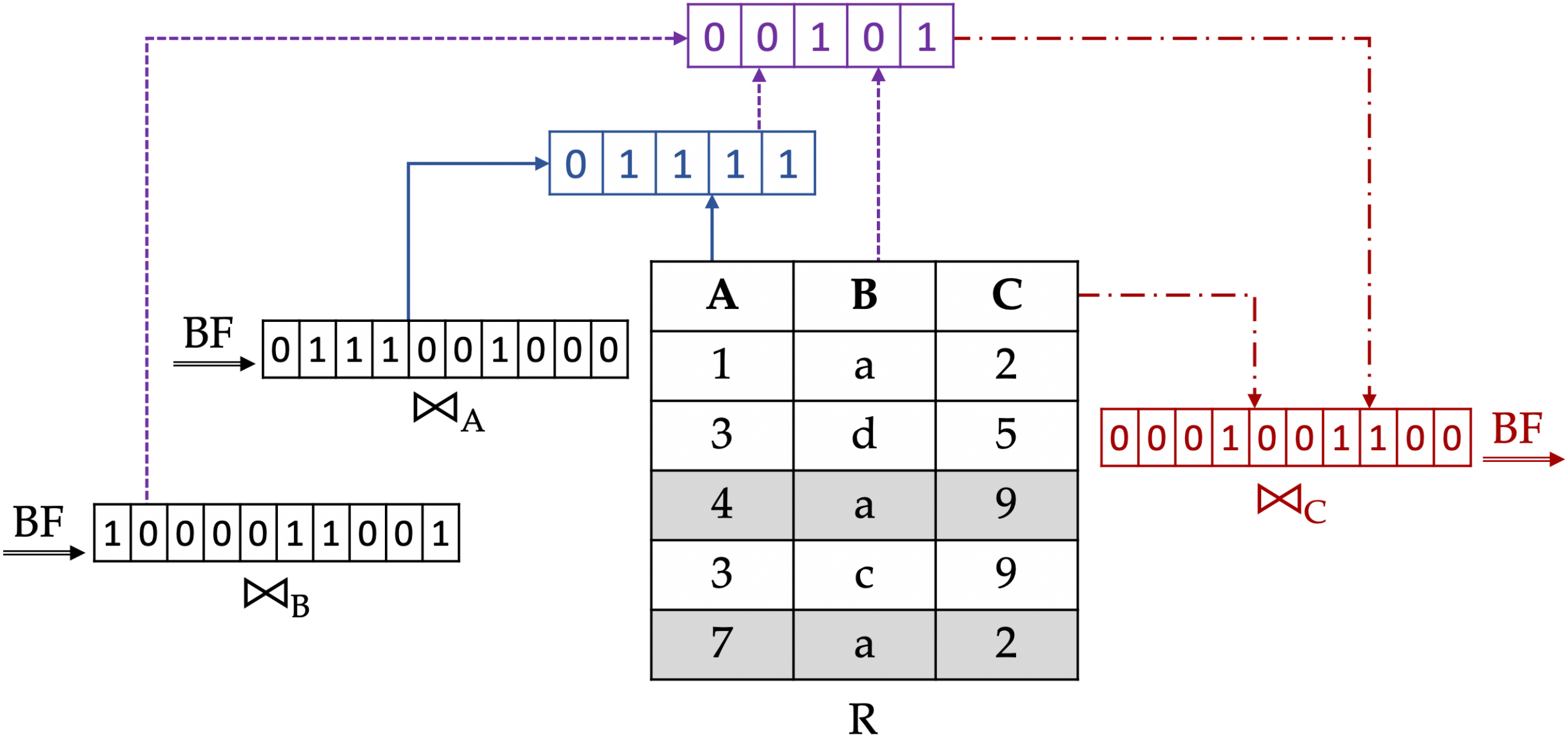}
    \caption{Filter Transformation \textnormal{--- Table $R$ receives two incoming filters on join attributes $A$ and $B$, and generates a transformed outgoing filter on join attribute $C$.}}
    \label{fig:filter-trans}
\end{figure}

Figure~\ref{fig:filter-trans} shows an example of filter transformation on table $R$. Table $R$ has three columns, and participates into three different joins on columns $A$, $B$, and $C$ respectively. In the predicate transfer phase, $R$ receives two incoming filters (assume we are using Bloom filters) on join attributes $A$ and $B$ respectively. Column $A$ is used to probe the incoming filter on join attribute $A$, where all rows except the first satisfy the join condition (with false positives). Then column $B$ on the remaining rows is used to probe the incoming filter on join attribute $B$, with another two rows (row 2 and 4) filtered out. Finally, column $C$ on the remaining two rows (row 3 and 5) is used to build the outgoing filter.

\vspace{0.05in}
\noindent
\textbf{Predicate Transfer Graph.} The join graph determines the topology of predicate transfer. Figure~\ref{fig:join-graph} shows the join graph for Query 5 in TPC-H. Each equi-join is represented as an edge. 
A \textit{predicate transfer graph} is a directed subgraph of the join graph. Transfers happen along the selected edges in the subgraph---local predicates of the source vertex are transferred to the target vertex as a filter. Figure~\ref{fig:pred-trans-graph} shows one predicate transfer graph of TPC-H Q5.

The topology of the predicate transfer graph affects the performance of the predicate transfer phase and also the selectivity of the transferred filters. In this paper, we use a simple heuristic that points an edge from a smaller table to a bigger table. The intuition is the same as why Bloom join builds Bloom filter at the smaller table---to reduce Bloom filter size and increase filter selectivity.  Our current heuristic does not remove any edge in the join graph when generating the predicate transfer graph. It also guarantees that the resulting graph is a Directed Acyclic Graph (DAG).
The predicate transfer graph in Figure~\ref{fig:pred-trans-graph} follows this heuristic.


\vspace{0.05in}
\noindent
\textbf{Transfer Schedule.} The transfer schedule determines when and how the predicates are transferred across the predicate transfer graph. Numerous design decisions can be made in the schedule. In particular, the schedule specifies which tables in the query should construct initial local filters to start the transfer process, and the order of issuing the remaining transfers. For each table that sends the local filter out, the schedule determines when the transfer happens---multiple transfers may happen in serial or parallel. Moreover, the transfer can happen back and forth, following both directions of certain edges. 
Pruning may be adopted to avoid non-beneficial transfers, and the transfer direction may be dynamically adjusted at runtime. Identifying a good transfer schedule is critical to the system performance. 


In this paper, we adopt a heuristic that builds the transfer schedule using one \textit{forward pass} and one \textit{backward pass} similar to the Yannakakis algorithm. 
The predicate transfer graph is determined at planning time and remains fixed during runtime. 
In the forward pass, we build initial local filters on the leaf nodes in the predicate transfer graph (i.e., nodes with only outgoing edges but no incoming edge). These filters are transferred following the topological order of the predicate transfer graph, which exists because the graph is a DAG. 
If one node has one or more incoming edges, the node will collect all the incoming filters before performing the transformation to produce outgoing filters (LIP-style~\cite{lip17} incoming filter ordering can be utilized for further optimization); the transformation will scan the table only once, regardless of the number of incoming or outgoing edges. 
The forward pass finishes once all filters are fully transferred. 

The system then starts the backward pass, where we simply reverse the direction of all edges and repeat the same process in the forward pass. After both passes are done, each table has been reduced based on the transformed filters it receives. The later join phase will start from these pre-filtered tables.

\begin{figure}[h]
\subfloat[Forward Pass]{
	\centering
	\includegraphics[width=0.8\columnwidth]{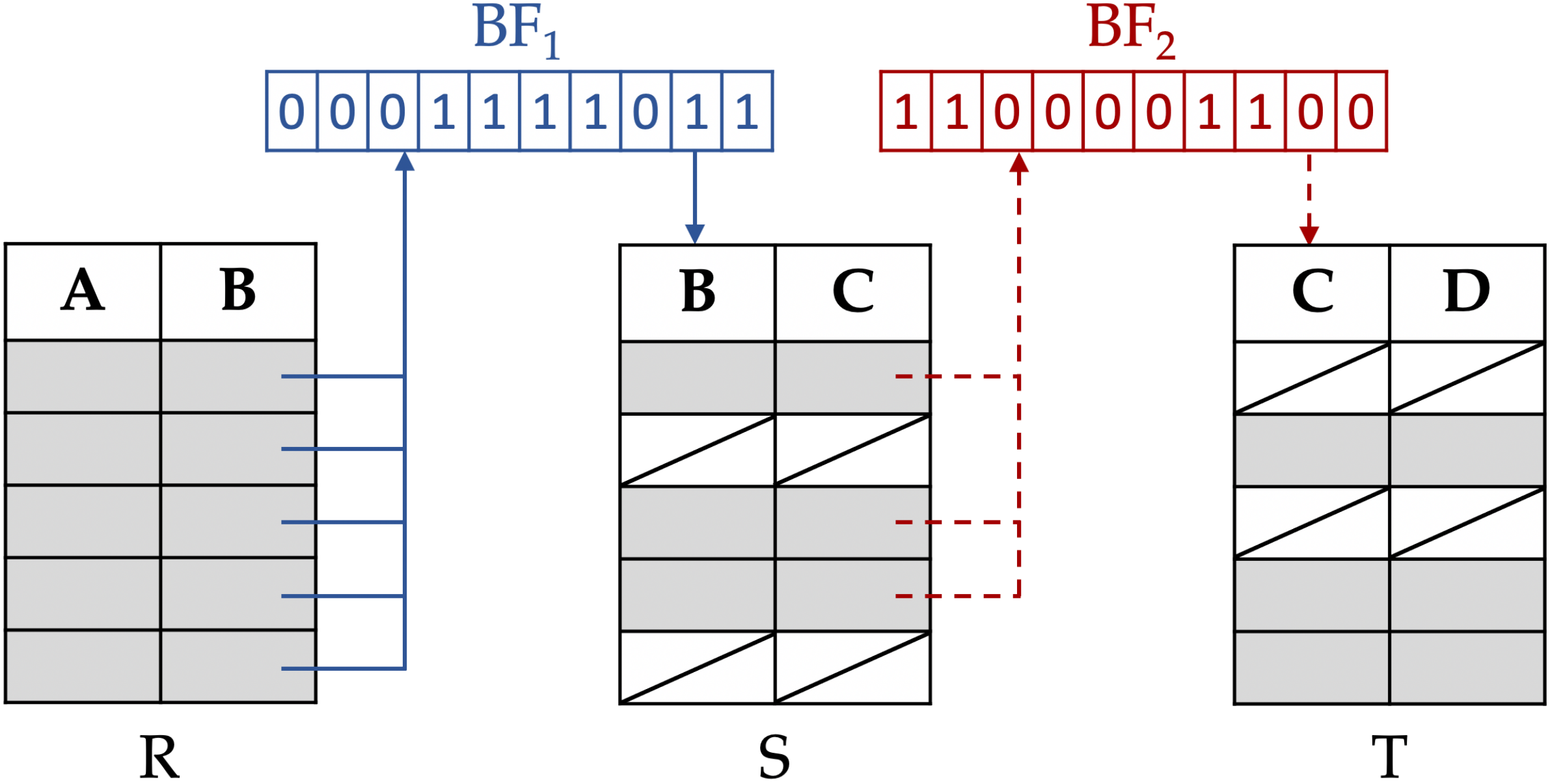}
	\label{fig:ex-forward}
} \\
\subfloat[Backward Pass]{
	\centering
	\includegraphics[width=0.8\columnwidth]{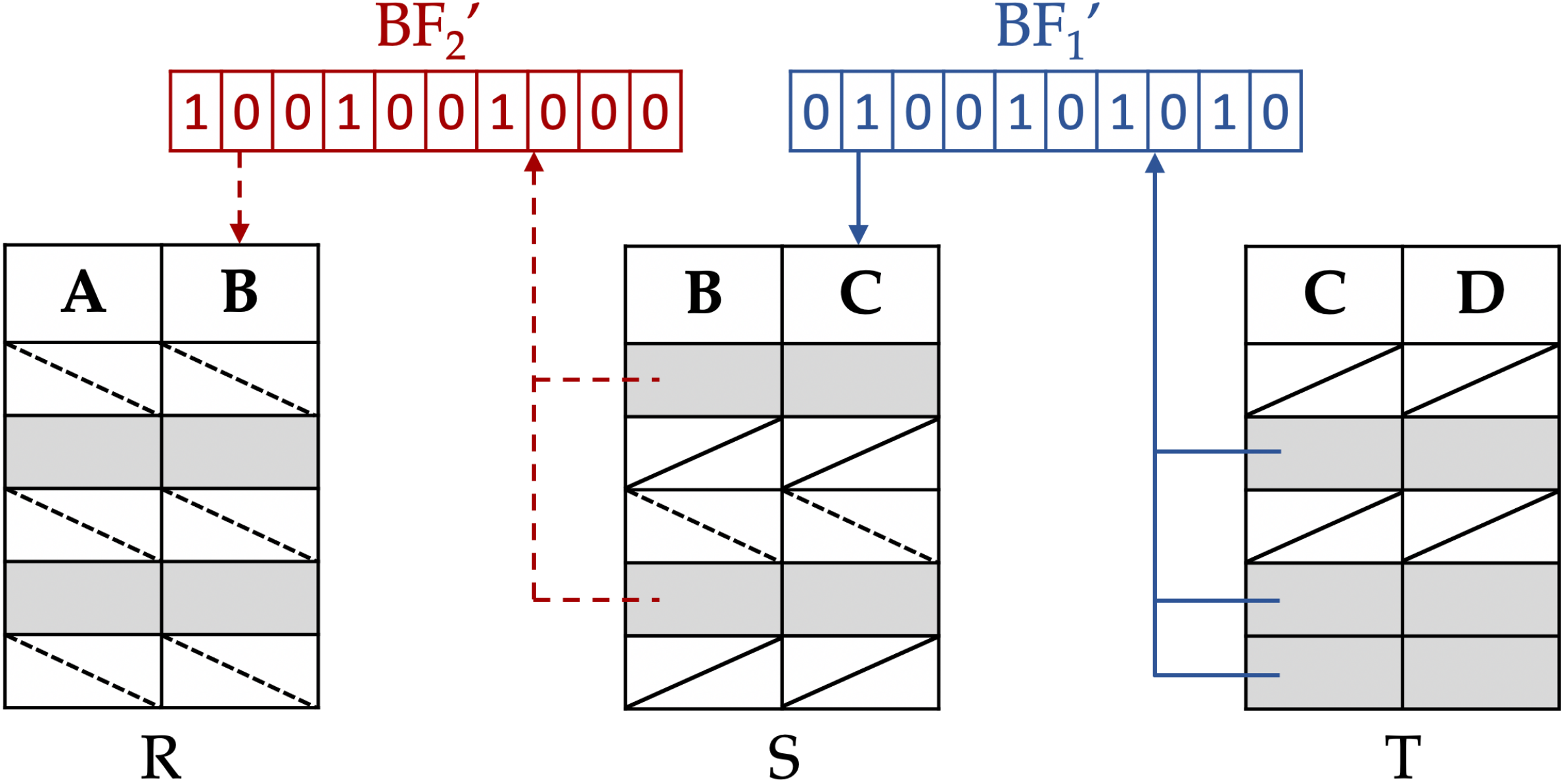}
	\label{fig:ex-backward}
}
\caption{Example of Predicate Transfer on a Join Query \textnormal{--- $R \bowtie S \bowtie T$.}}
\label{fig:ex}
\end{figure}

Figure~\ref{fig:ex} presents an example of predicate transfer on a join query, which joins three tables $R$, $S$, and $T$. Assume the transfer starts at table $R$ and all inherit local filters are already consumed. In the forward pass (Figure~\ref{fig:ex-forward}), $R$ first constructs a Bloom filter ($BF_1$) on join attribute $B$ and sends it to $S$. $S$ applies $BF_1$ to its join attribute $B$, which filters out two rows (row 2 and 5), and the remaining three rows are used to construct a new Bloom filter ($BF_2$) on another join attribute $C$. When $T$ receives the transformed filter from $S$, the join column $C$ is used to probe the filter where row 1 and 3 are removed. The backward pass (Figure~\ref{fig:ex-backward}) starts with input as the remaining rows of the forward pass. Table $T$ constructs a Bloom filter ($BF_1'$) on join attribute $C$, which is applied on $S.C$, with row 3 filtered out. Then $S$ constructs another Bloom filter ($BF_2'$) on the join column $B$, which is able to filter three rows (row 1, 3, and 5) from $R$.
    
A more complicated example is TPC-H Q5, which is shown in Figure~\ref{fig:pred-trans-graph}. The first Bloom filter is constructed on \texttt{region}, and sent to \texttt{nation}. The filter is then transformed into two outgoing filters which are sent to \texttt{customer} and \texttt{supplier} respectively. Similarly, \texttt{supplier} transfers two outgoing filters following the edges to \texttt{customer} and \texttt{lineitem}. At \texttt{customer}, two separate incoming filters are applied with one outgoing filter produced and sent to \texttt{order}, which is then transformed and sent to \texttt{lineitem}. The forward pass finishes when both incoming filters arrive at \texttt{lineitem}, and after that the backward pass begins in a symmetric way.



\vspace{0.05in}
\noindent
\textbf{Filter Type.} Our discussion so far uses Bloom filters to represent the predicates. In fact, other representation of filters can also be used. If a precise representation is used, i.e., the filter precisely encodes all the join keys, then a transfer becomes a semijoin and the algorithm becomes similar to Yannakakis. An ideal filter should be efficient to construct and check, and achieve low false positive rates. We use Bloom filters in our prototype since it is the best candidate available today. But predicate transfer can automatically benefit from any potential improvement in filtering techniques.

\vspace{0.05in}
\noindent
\textbf{Transfer Path Pruning.} As discussed above, our current scheduling heuristic makes two full passes of the predicate transfer graph. In practice, some transfers may not increase filter selectivity but consume computational resources. 
An intelligent transfer scheduler should identify such scenarios and stop transferring these filters further to avoid wasting CPU cycles. Such transfer path pruning can be done at either planning time or runtime. Our current prototype does not incorporate any pruning and always performs the forward and backward passes in full. We observe this already demonstrates significant performance improvement, and believe incorporating path pruning will lead to even larger speedups. 


\subsection{Join Phase}
\label{sec:join-phase}

After the predicate transfer phase completes, each table may have already been processed by several filters, including the inherent filters from the query and the transferred filters. The join phase basically executes the original query with the reduced input tables.

\vspace{0.05in}
\noindent
\textbf{Unified Query Plan.} As a straightforward design, the database can directly execute the query plan as a regular query in the join phase, with the leaf nodes (i.e. scan) replaced by the filtered tables produced by the predicate transfer phase. The predicate transfer schedule is essentially also a query plan. The two query plans can be concatenated such that the leaf nodes in the join plan are just the output nodes of the predicate transfer schedule. This avoids rescanning in the join phase and requires no changes to the executor---the executor is oblivious to the predicate transfer phase and executes the modified query plan regularly.

\vspace{0.05in}
\noindent
\textbf{More Accurate Cardinality Estimation.} The predicate transfer phase updates the cardinality of the input tables in the join phase. Therefore, the original query plan generated beforehand may become suboptimal based on the stale cardinalities. A replanning step between the two phases may produce a better plan that leads to further performance improvement. Although join performance will be more robust to join orders (as will be shown in Section~\ref{sec:eval-q5}), performance can still be affected by the quality of the query plan, with the factors including the size of materialized intermediate tables, which table to build the hash table and which table to probe, etc. Moreover, similar to the Yannakakis algorithm, predicate transfer bounds the size of the intermediate join tables in the join phase (Section~\ref{sec:discuss}), which can be utilized to improve cardinality estimation.

\subsection{Extension to General Queries}

In the discussion above, we assume table joins are inner equi-joins, and cover queries with only joins and local filters (filters over base tables). In this section, we extend the predicate transfer mechanism to further support general queries. 


\vspace{0.05in}
\noindent
\textbf{Supporting More Operators in Predicate Transfer Graph.} 
We first extend predicate transfer to support outer joins. In particular, a left outer join operation can be incorporated into the predicate transfer graph by allowing predicate transfer in only one direction, i.e., from the left table to the right table; but the other transfer direction is blocked. Therefore, such a transfer can happen in either forward pass or backward pass, but not in both passes. 
A right outer join can be supported in a similar way. A full outer join, however, cannot be incorporated into the predicate transfer graph. 

Considering more general operators, we note that an operator will block predicate transfer if it does not preserve the join key during the computation (e.g., perform scalar aggregations on the join key). 
In particular, we identify the following operators that can also be incorporated into the predicate transfer graph.

\begin{itemize}
    \item Operators including filters between intermediate join tables, column projection, sorting, and top-K do not block predicate transfer.
    \item Grouped aggregation does not block predicate transfer when the join key is a subset of the group key.
    \item Scalar user-defined functions do not block the transfer to the downstream join, but may block the transfer to the upstream join if the function is not invertible.
\end{itemize}

\vspace{0.05in}
\noindent
\textbf{Beyond a Single Predicate Transfer Graph.} 
Some queries may contain operators that cannot be incorporated into a predicate transfer graph. Example operators include but are not limited to full outer joins, scalar aggregations, and group-by aggregations where the join key is being aggregated. When such a scenario is encountered, we can apply predicate transfer only on a subset or several subsets of the query execution plan, and use conventional methods to execute the rest of the query. For example, this means a query can be partially executed first, leading to a subquery plan that can be represented as a predicate transfer graph in order to apply predicate transfer. After the predicate transfer phase and the join phase, the rest of the query can continue execution. It is also possible that predicate transfer can be applied multiple times to different parts of the query plan---the predicate transfer phase and regular query execution can alternate. 


In our current prototype, we apply the heuristic that first identifies and executes single-table subquery plans (e.g., group by aggregation on a single table) before the predicate transfer phase. 

\subsection{Cost Analysis}
\label{sec:discuss}

Compared to the Yannakakis algorithm, predicate transfer does not provide theoretical optimality, but it is more versatile. Predicate transfer supports both precise filters (like semi-join) and Bloom filters, any join-graph topology, outer joins and cyclic queries, more operators, and complex predicate transfer schedules. 


In this section, we present a simple cost analysis of predicate transfer compared to the Yannakakis algorithm and show that predicate transfer is more efficient and robust than Yannakakis, and can achieve close to optimal pre-filtering efficiency. Our key idea is to show that using the cheap Bloom filters drastically reduces the cost of excessive hash probes in the semi-join phase of Yannakakis, filters out most tuples not participating in the joins, and only incurs bounded amount of false positives to be removed in the join phase.

\vspace{0.05in}
\noindent
\textbf{Cost Model.}
Let $t$ be the number of tables in a given join query and $N$ be the input size (i.e. the total number of tuples in all joining tables). We assign a unit cost to each per-tuple scan, hash table insertion or probe, and a $\beta$ cost per-tuple for Bloom filter insertion or probe. As a Bloom filter is of a small size and thus likely to be cache resident, Bloom filter operations are typically much cheaper than hash table operations, i.e. $\beta \ll 1$. We assume that the Bloom filter has a false positive rate of $\epsilon \ll 1$ that can be appropriately configured (e.g., we can tune $\epsilon$  to be smaller by increasing the Bloom filter size or number of hash functions, but this makes $\beta$ larger). The reader can refer to~\cite{lip17} for an in-depth study on Bloom filter configurations. 

\vspace{0.05in}
\noindent
\textbf{Yannakakis algorithm.}
At the semi-join phase, scanning tables to build or probe hash tables cost $N$ units, independent of the direction of the forward/backward semi-join passes. The cost of building or probing intermediate hash tables can be bounded by $c_y \cdot N$, where $c_y$ is a constant highly sensitive to the choice of the rooted join tree of the query. An ideal join tree and orientation 
can drastically reduce the size of intermediate hash tables, leading to a cheaper semi-join phase (smaller $c_y$). The join phase of Yannakakis is perfectly robust, as every join order costs $t \cdot \mathsf{OUT}$ units of hash table accesses. 

\vspace{0.05in}
\noindent
\textbf{Predicate Transfer.}
At the predicate transfer phase, scanning tables to build or probe Bloom filters costs $N$ units. As we only build or probe Bloom filters, the cost can be bounded by $\beta \cdot c_p \cdot N$ units, where $c_p$ is a constant that depends on the predicate transfer graph topology and the transfer schedule. As $\beta \ll 1$, the sensitivity of the runtime to the constant $c_p$ shrinks by a factor of $\beta$.

Let $T_k$, $T^*_k$ be the size of the $k$th join table before and after a semi-join phase of Yannakakis. The predicate transfer phase, however, passes a larger table of size $T^*_k + (T_k - T^*_k) \cdot \epsilon$ to the join phase, by an approximated factor of
$$
p = \prod_{k=1}^{t} \left (1 + \frac{T_k - T^*_k}{T^*_k} \cdot \epsilon \right).
$$
Assume 
$$
\hat{k} = \argmax_k \frac{T_k - T^*_k}{T^*_k},
$$
then
$$
p \leq \left (1 + \frac{T_{\hat{k}} - T^*_{\hat{k}}}{T^*_{\hat{k}}} \cdot \epsilon \right)^t = (1 + \epsilon')^t \approx 1 + \epsilon' t,
$$
ignoring higher order terms of $\epsilon'$, and assuming
$$
\epsilon' = \frac{T_{\hat{k}} - T^*_{\hat{k}}}{T^*_{\hat{k}}} \epsilon = \left (\frac{1}{Sel_{\hat{k}}} - 1 \right) \epsilon \ll 1,
$$
where $Sel_{\hat{k}}$ denotes the smallest join selectivity in the joining tables.

Then in the join phase, the cost of predicate transfer can be approximated as $t \cdot \mathsf{OUT} \cdot (1 + \epsilon' t) $ units. The choice of the join order only affects the extra $\epsilon' t^2 \cdot \mathsf{OUT}$ term. Under our assumption that $\epsilon' \ll 1$ (and thus $\epsilon' t^2 \cdot \mathsf{OUT}$ is small), the join phase still attains promising robustness.

As a summary, the Yannakakis algorithm guarantees maximum filtering at the semi-join phase and perfect robustness at the join phase, but at the cost of a much more expensive and unstable semi-join phase (our evaluation in Section~\ref{sec:eval-q5} verifies this).
In contrast, predicate transfer addresses the shortcomings via a more stable and efficient Bloom filter transfer scheme,
while maintaining near-maximum filtering capabilities at the predicate transfer phase and near-perfect robustness in the join phase. 









\section{Evaluation}
\label{sec:eval}

This section presents our preliminary evaluation results. We describe the experimental setup in Section~\ref{sec:eval-setup} and compare predicate transfer with baseline join strategies in Section~\ref{sec:eval-main-results}. 
Then we perform a deep drive to understand the performance on TPC-H Q5 in Section~\ref{sec:eval-q5}.

\subsection{Experimental Setup}
\label{sec:eval-setup}

We conduct experiments on a single AWS EC2 r5.4xlarge instance, with 16vCPU and 128GB memory. The server runs the Ubuntu 20.04 operating system. We use the widely adopted data analytics benchmark, TPC-H, with 22 queries in total. We use both an 1GB data set (a scale factor of 1) and a 10GB data set (a scale factor of 10). Queries are executed on a single CPU core. For all the experiments, we measure the in-memory query performance by running the query twice, where the first run loads all the tables into the memory, and we measure the performance of the second run.

The testbed we use on evaluation is FlexPushdownDB~\cite{fpdb} (\textit{FPDB} in short), an open-source cloud-native OLAP DBMS. Table data is placed on local disks in Parquet~\cite{parquet} format unsharded. FPDB leverages join and Bloom filter implementation of Apache Arrow~\cite{arrow}. The evaluation results may vary on different DBMSs, depending on the performance ratio between the join and Bloom filter implemented.


We compare the proposed join strategy \pt with three other baselines: \nopt, \bj, and \yk. \nopt does not transfer predicates among joining tables---pairs of tables are joined regularly as in most DBMSs. \bj performs one-hop predicate transfer between joining table pairs, where the build side constructs a Bloom filter which is used to filter the probe side. \yk executes the semi-join phase of the Yanakakis algorithm ahead of the join phase.

Since the vanilla Yannakakis algorithm is only applicable on acyclic conjunctive queries, we make two extensions to make it applicable on all TPC-H queries. First, we adopt the same mechanisms that \pt deploys to handle the case of outer joins and non-join operators in the query plan. Second, for cyclic queries like Q5 and Q9, we break the cycle in the join graph by randomly picking a root node and, performing a BFS search from the root. The result join tree represents the transfer order of the semi-join phase in \yk. 

\subsection{TPC-H Performance}
\label{sec:eval-main-results}

\begin{figure*}[ht]
    \subfloat[SF = 1]{
        \includegraphics[width=\linewidth]{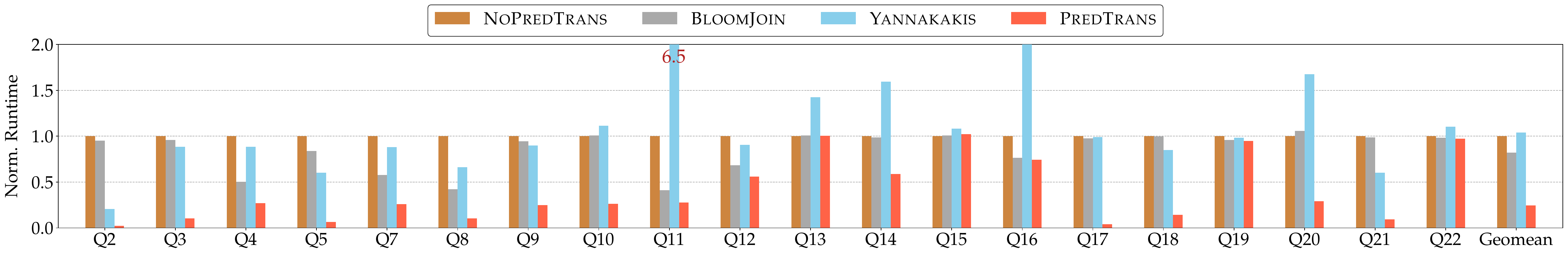}               
    }
    \\
    \subfloat[SF = 10]{
        \includegraphics[width=\linewidth]{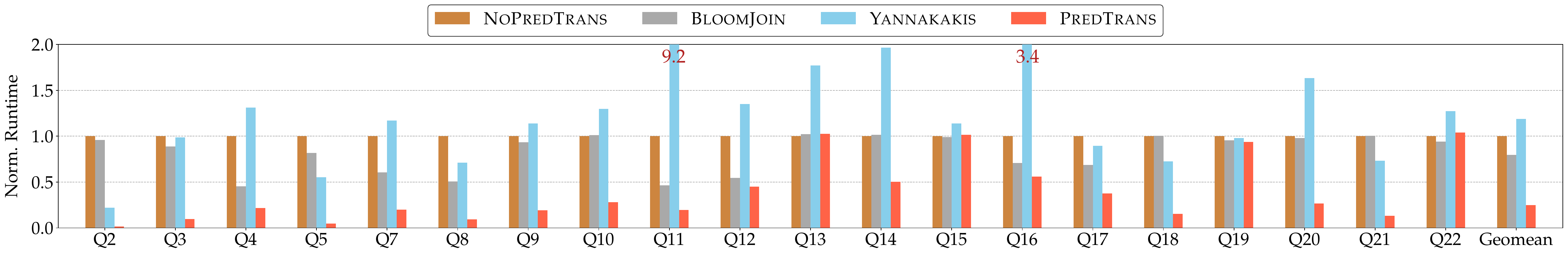}                
    }
    \caption{Performance Evaluation of Predicate Transfer on TPC-H \textnormal{(normalized to \nopt)}.}
    \label{fig:exp-tpch}
\end{figure*}

Figure~\ref{fig:exp-tpch} shows the execution time of different predicate transfer strategies on TPC-H queries. Since Q1 and Q6 involve no joins, we exclude them from the benchmark. On average, \pt outperforms \nopt by 4.1$\times$, \bj by 3.3$\times$, and \yk by 4.2$\times$ for SF1, and 4.0$\times$, 3.2$\times$, and 4.8$\times$ for SF10, respectively. 15 queries out of 20 see performance improvement on both data sets.

\pt achieves significant performance improvement on queries with a large amount of joins. Half of the queries include joins across at least four tables. Among this, Q2 (joins across nine tables) benefits most from predicate transfer, which outperforms \nopt and \bj by 47$\times$ and 44$\times$ on SF1, as well as 63$\times$ and 61$\times$ on SF10, respectively. Through predicate transfer, filter predicates on tables \texttt{Part} and \texttt{Region} are sent to every other table in the join graph through lightweight Bloom filters. As a result, the predicate transfer phase of \pt reduces the size of input tables that participate in the join phase by over 99\%, such that the expensive join operations are only performed on a tiny fraction of data. The \yk algorithm outperforms both \nopt and \bj baselines by over 4$\times$ on both data sets. In fact, compared to \pt, \yk can pre-filter even more unnecessary data records ahead of the join phase since the Bloom filters leveraged by \pt incur false positives. However, the small performance benefit within the join phase is overwhelmed by the large overhead brought by semi-joins, making \yk perform worse that \pt.

On SF1, we observe the highest speed up on queries Q2, Q5, Q17, Q18, and Q21, between 7$\times$ to over 44$\times$. In these queries, there is a subquery executed with the results joined with the tables in the main query, and the large fact table are accessed by both the main query and the subquery (e.g., \texttt{Lineitem} in Q17 and Q18). Since \nopt and \bj perform no predicate transfer and one-hop transfer only, a single filter predicate cannot be sent to both the main query and the subquery to pre-filter the corresponding fact table. Conversely, \pt broadcasts every filter predicate globally inside the join graph, such that both fact tables in the main query and subquery can be filtered. Moreover, Q17 joins base tables with aggregation results. By executing the aggregation beforehand, predicate transfer is able to achieve a higher selectivity by starting transfer from a smaller intermediate result table.
On SF10, Q2 and Q5 achieve more improvement on \pt than SF1. The speedup on Q18 is similar on both data sets. However, for query Q17 and Q21, compared to SF1, the speedup on SF10 of \pt shrinks --- \pt outperforms \bj by 1.8$\times$ on Q17 and 7.6$\times$ on Q21. On Q17, the grouped aggregation on the large fact table dominates the execution time, where the pre-filtering cannot make significant impact. Q21 involves a large amount of joins, which amplifies the amount of false positives brought by Bloom filters. As a result, the amount of remaining rows after the transfer phase is still about 2$\times$ as the amount of rows that actually contribute to the final join output.

Queries with fewer join operations benefit less from predicate transfer (e.g., Q13, Q14), since one-hop predicate transfer may already be enough to forward local filter predicates to the global. However, we still observe a large speedup on Q3 (over 9$\times$ on both SF1 and SF10). Q3 joins three relatively large tables \texttt{customer}, \texttt{orders}, and \texttt{lineitem}. Since all three tables have local filters, \bj can only transfer a portion of them within a single hop. Instead, \pt can make sure each table receives the transformed filter predicates of every other table, which maximizes the effectiveness of the pre-filter phase.

Another interesting observation is that \yk may not always outperform \nopt and \bj baselines. For example, \yk underperforms \bj by 16$\times$ and 20$\times$ on Q11 on both data sets respectively, and by 4.8$\times$ on Q16 with SF10. One cause is that the Yannakakis algorithm does not specify the root of the join tree in the semi-join phase, and a bad semi-join order may construct several large hash tables at the beginning. However, this is not an issue in \pt since we use a heuristic to transfer from smaller tables to larger tables (see Section~\ref{sec:transfer-phase}), minimizing the memory stalls incurred by bitmap operations.

We further identify queries that have operators that block predicate transfer (e.g. outer joins), which are Q13, Q15, Q16, and Q22. Q13 and Q16 have left or right outer joins which block predicate transfer on one direction, Q15 has a join where one input is the result of scalar aggregation, and Q22 contains blocking operators of both kinds. The speedup on these queries of \pt is less than other queries, since the transfer is restricted within a small portion of the join graph (e.g. between only two tables), which weakens the pre-filtering power in the predicate transfer phase.

\subsection{Case Study --- TPC-H Q5}
\label{sec:eval-q5}

To get a deeper understanding of the performance benefits, we conduct a detailed analysis on Q5, one of the complex queries in TPC-H. The query performs inner joins across six tables, and the join graph is shown in Figure~\ref{fig:join-graph}.

\begin{table}[h]
  \caption{Join Table Size in Q5 (SF = 1) \textnormal{---\textit{HT} denotes the number of rows in the hash table, and \textit{PR} represents the number of rows that probe the hash table.}}
  \label{tab:tbl-size-sf1}
  \footnotesize
  \begin{tblr}{
    colspec = {c|cc|cc|cc|cc},
    cell{3}{8-9} = {green!15},
    cell{4}{8-9} = {green!15},
    cell{5}{8-9} = {green!15},
    cell{6}{8-9} = {green!15},
    cell{7}{8-9} = {green!15},
  }
    \toprule
    & \SetCell[c=2]{c} {\sc \textbf{NoPredTrans}} & & \SetCell[c=2]{c} {\sc \textbf{BloomJoin}} & & \SetCell[c=2]{c} {\sc \textbf{Yannakakis}} & & \SetCell[c=2]{c} {\sc \textbf{PredTrans}}\\
    \cmidrule{2-9}
    & \textbf{HT} & \textbf{PR} & \textbf{HT} & \textbf{PR} & \textbf{HT} & \textbf{PR} & \textbf{HT} & \textbf{PR}\\
    \cmidrule{1-9}\morecmidrules\cmidrule{1-9}
    \textbf{Join 1} & 10K & 6M & 10K & 6M & 2K & 181K & 2K & 74K\\
    \cmidrule{1-9}
    \textbf{Join 2} & 228K & 6M & 228K & 1M & 133K & 181K & 44K & 67K\\
    \cmidrule{1-9}
    \textbf{Join 3} & 150K & 910K & 150K & 44K & 69K & 193K & 15K & 60K\\
    \cmidrule{1-9}
    \textbf{Join 4} & 25 & 36K & 25 & 36K & 5 & 8K & 5 & 7K\\
    \cmidrule{1-9}
    \textbf{Join 5} & 1 & 36K & 1 & 7K & 1 & 8K & 1 & 7K\\
    \bottomrule
  \end{tblr}
\end{table}

\begin{table}[h]
  \caption{Join Table Size in Q5 (SF = 10) \textnormal{---\textit{HT} denotes the number of rows in the hash table, and \textit{PR} represents the number of rows that probe the hash table.}}
  \label{tab:tbl-size-sf10}
  \footnotesize
  \begin{tblr}{
    colspec = {c|cc|cc|cc|cc},
    cell{3}{8-9} = {green!15},
    cell{4}{8-9} = {green!15},
    cell{5}{8-9} = {green!15},
    cell{6}{8-9} = {green!15},
    cell{7}{8-9} = {green!15},
  }
    \toprule
    & \SetCell[c=2]{c} {\sc \textbf{NoPredTrans}} & & \SetCell[c=2]{c} {\sc \textbf{BloomJoin}} & & \SetCell[c=2]{c} {\sc \textbf{Yannakakis}} & & \SetCell[c=2]{c} {\sc \textbf{PredTrans}}\\
    \cmidrule{2-9}
    & \textbf{HT} & \textbf{PR} & \textbf{HT} & \textbf{PR} & \textbf{HT} & \textbf{PR} & \textbf{HT} & \textbf{PR}\\
    \cmidrule{1-9}\morecmidrules\cmidrule{1-9}
    \textbf{Join 1} & 100K & 60M & 100K & 60M & 20K & 1.8M & 20K & 835K\\
    \cmidrule{1-9}
    \textbf{Join 2} & 2.3M & 60M & 2.3M & 9.3M & 1.2M & 1.8M & 305K & 659K\\
    \cmidrule{1-9}
    \textbf{Join 3} & 1.5M & 9.1M & 1.5M & 499K & 689K & 1.9M & 150K & 376K\\
    \cmidrule{1-9}
    \textbf{Join 4} & 25 & 364K & 25 & 364K & 5 & 75K & 5 & 73K\\
    \cmidrule{1-9}
    \textbf{Join 5} & 1 & 364K & 1 & 73K & 1 & 75K & 1 & 73K\\
    \bottomrule
  \end{tblr}
\end{table}

\vspace{0.05in}
\noindent
\textbf{Join Table Size.} We measure the sizes of both input tables of each join, following the join order specified in the query plan (FPDB utilizes Apache Calcite~\cite{calcite} for query optimization like join ordering), and the results are shown in Table~\ref{tab:tbl-size-sf1} and Table~\ref{tab:tbl-size-sf10}.
\pt reduces the join table size by 98\% over \nopt, 96\% over \bj, and 64\% over \yk on SF1 (Table~\ref{tab:tbl-size-sf1}), and 98\% over \nopt, 92\% over \bj, and 67\% over \yk on SF10 (Table~\ref{tab:tbl-size-sf10}). In \bj, the largest fact table \texttt{lineitem} can only be pre-filtered after the first join, where the inner table \texttt{Orders} owns local filter predicates that can be trasferred to \texttt{lineitem}. \pt shows the superiority to be able to pre-filter all join tables ahead of the entire join phase.

We observe a higher selectivity in the predicate transfer phase achieved by \pt, compared to \yk. This is because \yk can only guarantee the optimal pre-filtering on acyclic queries. For a cyclic query (like Q5), some edges in cycles are removed to form a tree, 
which sacrifices the overall filtering power. 
Instead, the heuristic adopted by \pt allows us to perform transfer for every join (in Q5 there is no blocking operator) regardless the cyclicity of the join graph, resulting in more base table records pre-filtered ahead of the join phase. 

\begin{figure}[h]
\begin{minipage}{0.48\linewidth}
    \centering
    \includegraphics[width=\linewidth]{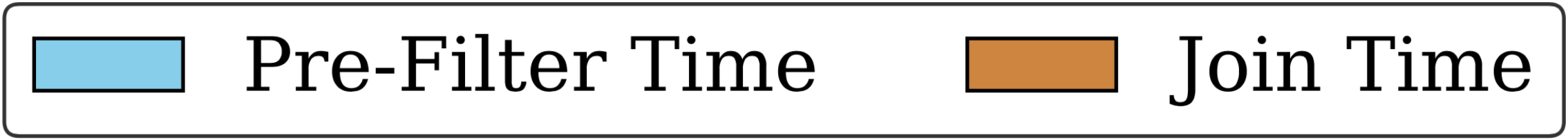}
\end{minipage}\\
\vspace{-0.15in}
\begin{minipage}{\linewidth}
    \subfloat[SF = 1]{
        \includegraphics[width=0.48\linewidth]{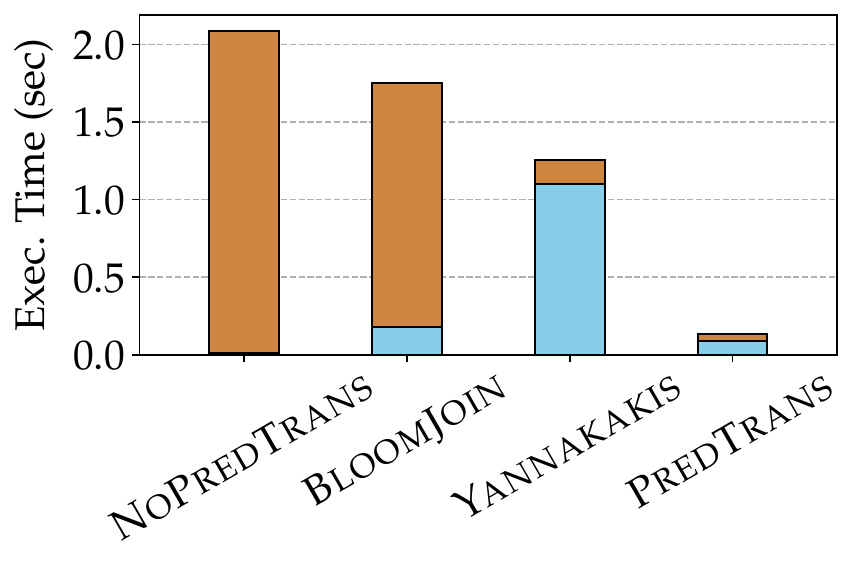}
    }
    \hspace{-0.1cm}
    \subfloat[SF = 10]{
        \includegraphics[width=0.48\linewidth]{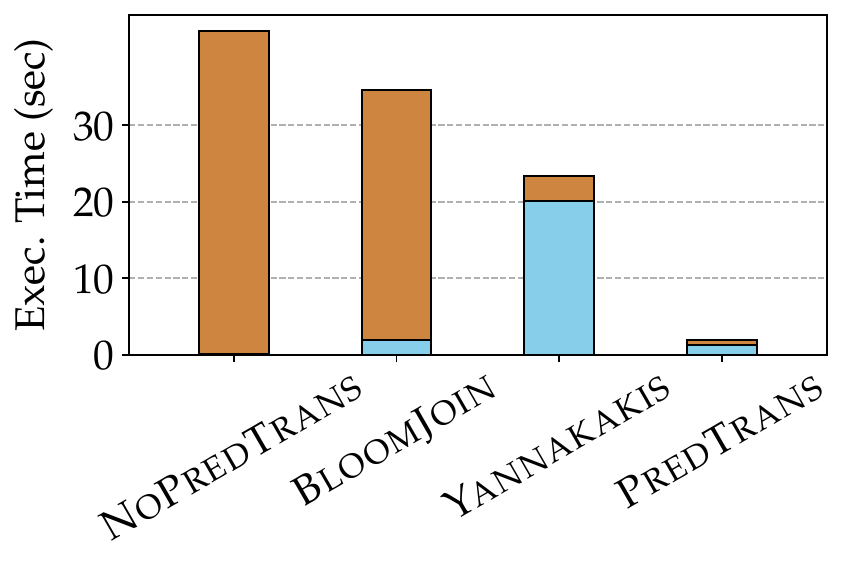}
    }
\end{minipage}
\caption{Performance Breakdown on TPC-H Q5.}
\label{fig:exp-break}
\end{figure}

\vspace{0.05in}
\noindent
\textbf{Performance Breakdown.} Figure~\ref{fig:exp-break} demonstrates the performance breakdown of Q5 in different predicate transfer strategies. The execution time is divided into pre-filter time and join time. Compared to \nopt and \bj, joins are accelerated by 44$\times$ and 34$\times$ in \pt on SF1, as well as 60$\times$ and 46$\times$ in SF10, due to the significant size reduction of the input join tables. \yk is also able to achieve a shrinkage on the input join tables, but the pre-filtering power is 3.3$\times$ and 4.7$\times$ less than \pt on SF1 and SF10 respectively since not all edges in the join graph are traversed. Moreover, the semi-joins it relies on are computationally expensive and dominate the entire execution time. As a result, the predicate transfer phase in \pt outperforms the semi-join phase in \yk by 13$\times$ and 16$\times$ on both data sets respectively, since bit operations used in Bloom filters are much cheaper than the construction and probe of the hash tables.

\begin{figure}[h]
\begin{minipage}{0.9\linewidth}
    \centering
    \includegraphics[width=\linewidth]{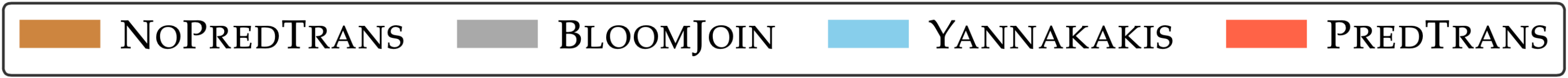}
\end{minipage}\\
\vspace{-0.1in}
\begin{minipage}{\linewidth}
    \subfloat[SF = 1]{
        \includegraphics[width=0.48\linewidth]{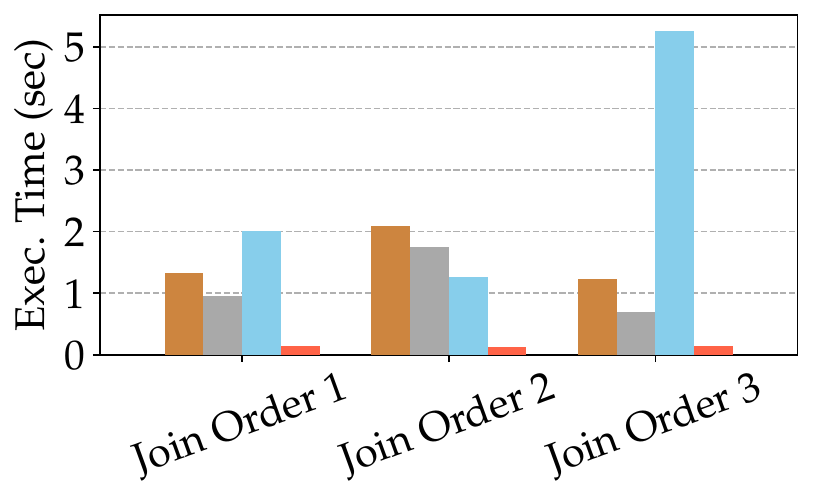}
    }
    \hspace{-0.1cm}
    \subfloat[SF = 10]{
        \includegraphics[width=0.48\linewidth]{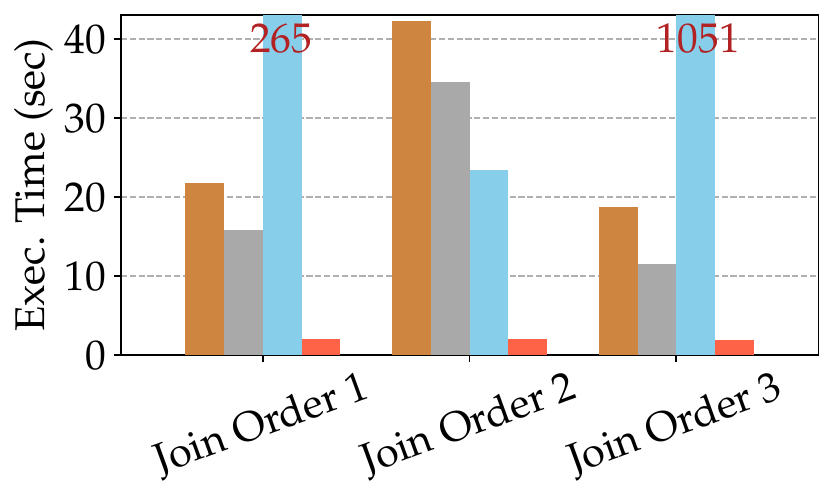}
    }
\end{minipage}
\caption{Performance of Different Join Orders on TPC-H Q5.}
\label{fig:exp-robust}
\end{figure}

\vspace{0.05in}
\noindent
\textbf{Robustness.} We next evaluate the sensitiveness on join orders for different predicate transfer strategies. We pick three different join orders and the result is shown in Figure~\ref{fig:exp-robust}. In both data sets, \pt achieves the best performance and outperforms other predicate transfer strategies on all the join orders. Notably, the the join order makes a much smaller performance variance in \pt (up to 12\%) compared to other strategies (up to 45$\times$). We further notice that for just the join phase, \yk achieves the similar robustness as \pt does. Both \yk and \pt is able to bound the size of the intermediate join results, making their join phase robust to different join orders. The performance of the expensive semi-joins used in \yk is unstable since the join tree construction is not deterministic, making it overall highly sensitive to the join order. Conversely, the heuristic adopted in the predicate transfer phase of \pt always points to the same transfer topology and schedule regardless of the join order.

\section{Conclusion and Future Work}
\label{sec:conclusion}

A new join optimization, predicate transfer, is proposed in this paper. Inspired by Yannakakis algorithm, predicate transfer generalizes Bloom join to transfer table-local filters to pre-filter multiple other tables. We laid out the design space of predicate transfer and described the heuristics used in our prototype. Evaluation showed an average $3.3\times$ speedup over Bloom join on TPC-H benchmark. 

Predicate transfer opens up substantial research opportunities, including better heuristics in the predicate transfer schedule, deeper theoretical analysis on the performance guarantees, and extending the technique to parallel and distributed environment. Discussions of these topics are left as future work.

\small
\bibliographystyle{ACM-Reference-Format}
\bibliography{ref}

\end{document}